\documentclass[useAMS,usenatbib]{mn2e}
\usepackage{epsfig}

\def\mnras{MNRAS}  
\def\apj{ApJ}      
\def\apjl{ApJL}    
\def\apjs{ApJS}    
\def\aap{A\&A}     
\def\aj{AJ}        
\def\araa{ARAA}    
\def\nat{Nature}   

\title[Ram pressure stripping hot gaseous halos]{Ram 
pressure stripping the hot gaseous halos of galaxies in groups 
and clusters}

\author[I. G. McCarthy et al.]{I. G. McCarthy$^{1}$\thanks{E-mail:
i.g.mccarthy@durham.ac.uk (IGM)}, C. S. Frenk$^1$, A. S. 
Font$^1$, C. G. Lacey$^1$, R. G. Bower$^1$, \newauthor N. L. 
Mitchell$^1$, M. L. Balogh$^2$ and T. Theuns$^{1,3}$
\\
\\
$^{1}$Department of Physics, University of Durham, South Road,
Durham, DH1 3LE\\
$^{2}$Department of Physics and Astronomy, University of 
Waterloo, Waterloo, ON, N2L 3G1, Canada\\
$^{3}$Department of Physics, University of Antwerp, Campus
Groenenborger, Groenenborgerlaan 171, B-2020 Antwerp, Belgium
}

\begin{document}

\date{Accepted XXXX. Received XXXX; in original form XXXX}

\pagerange{\pageref{firstpage}--\pageref{lastpage}} \pubyear{2006}

\maketitle

\label{firstpage}

\begin{abstract}

We use a large suite of carefully controlled full hydrodynamic 
simulations to study the ram pressure stripping of the hot 
gaseous halos of galaxies as they 
fall into massive groups and clusters.  The sensitivity of the 
results to the orbit, total galaxy mass, and galaxy structural 
properties is explored.  For typical structural and orbital 
parameters, we find that $\sim 30\%$ of the initial hot 
galactic halo gas can remain in place after 10 Gyr. We propose a 
physically simple analytic model that describes the stripping 
seen in the simulations remarkably well.  The model is 
analogous to the original formulation of Gunn \& Gott (1972), 
except that it is appropriate for the case of a spherical (hot) 
gas distribution (as opposed to a face-on cold disk) and takes 
into account that stripping is not instantaneous but occurs 
on a characteristic timescale.  The model reproduces the 
results of the simulations to within $\approx 10\%$ at 
almost all times for all the orbits, mass ratios, and galaxy 
structural properties we have explored.  The one exception  
involves unlikely systems where the orbit of the galaxy is 
highly non-radial and its mass exceeds about 10\% of the group 
or cluster into which it is falling (in which case the model 
under-predicts the stripping following pericentric passage).  
The proposed model has 
several interesting applications, including modelling the ram 
pressure stripping of both observed and cosmologically-simulated 
galaxies and as a way to improve current semi-analytic models of 
galaxy formation.  One immediate consequence is that the colours 
and morphologies of satellite galaxies in groups and clusters 
will differ significantly from those predicted with the standard 
assumption of complete stripping of the hot coronae.

\end{abstract}


\section{Introduction}

There are marked differences in the observed properties of 
the field and cluster galaxy populations.  Perhaps the best known 
difference is the larger fraction of galaxies that are 
ellipticals or S$0$s (and the correspondingly lower spiral 
fraction) in clusters relative to the field (e.g., 
Dressler 1980; Goto et al.\ 2003).  Not only are the 
morphologies of cluster galaxies different from those of field 
galaxies, but so too are a variety of their other observed 
properties, including colours (e.g., Balogh et al.\ 2004; 
Hogg et al.\ 2004), star forming properties (e.g., 
Poggianti et al. 1999; Balogh et al.\ 2000; Gomez et al.\ 2003), 
and the distribution and total mass of their gaseous 
component (e.g., Cayatte et al.\ 1994; Solanes et al.\ 
2001).  These observed differences indicate that the dense 
environments of groups and clusters are somehow strongly 
modifying the properties of galaxies as they fall in.  

Uncovering the physical mechanisms that give rise to the observed 
variation in 
galaxy properties has been an active topic of research over the 
past two or three decades (e.g., Dressler 1984; Sarazin 1988).  
One of the most commonly mentioned processes is ram pressure 
stripping (Gunn \& Gott 1972).  Here the gaseous component 
(which can be composed of both cold atomic/molecular gas and 
a hot extended component) of the orbiting galaxy is subjected 
to a wind due to its motion relative to the 
intracluster medium (ICM).  The gas will be stripped if the wind 
is sufficiently strong to overcome the gravity of the galaxy. 
Recently, direct observational evidence for the ram pressure 
stripping of galaxies in clusters has been provided by long 
(up to tens of kpc) tails of gas seen to be trailing behind 
several cluster galaxies (e.g., Sakelliou et al.\ 2005; Crowl et 
al.\ 2005; Vollmer et al.\ 2005; Sun \& Vikhlinin 2005; Machacek 
et al.\ 2006; Sun et al.\ 2007a).  Such stripping could at least 
partially account for the differing properties of cluster and 
field galaxies.

There have been numerous theoretical studies dedicated to 
calculating the effects of ram pressure stripping on galaxies 
using hydrodynamic simulations or semi-analytic models.  
The vast majority of these studies have focused on the stripping 
of cold gaseous disks with an emphasis on whether this can 
account for the observed lower star formation rates (and redder 
colours) of cluster spirals relative to their field counterparts 
(e.g., Abadi et al.\ 1999; Quilis et al.\ 2000; Vollmer et al.\ 
2001; Okamoto \& Nagashima 2003; Mayer et al.\ 2006; Roediger 
et al.\ 2006; Hester 2006; Jachym et al.\ 2007; Roediger \& 
Br{\"u}ggen 2006; 2007).  However, the stripping of 
extended {\it hot} gaseous halos of galaxies is 
only just beginning to be explored (e.g., Kawata \& Mulchaey 
2007) and has not yet been studied in a detailed and systematic 
way.  The hot extended 
component is predicted to exist around most massive galaxies by 
semi-analytic models and cosmological simulations 
and is directly observable at X-ray wavelengths in the case of 
normal ellipticals.  If the hot gaseous halo is completely 
stripped (as is typically assumed), the only fuel 
available for star formation is that which resided in the cold 
component when the galaxy first fell into the cluster.   
(This process of removing the supply of halo gas is sometimes 
referred to as ``strangulation'' or ``starvation''.)  However, 
if the hot halo remains intact for some time it can, via 
radiative cooling losses, replenish the cold component and 
potentially significantly prolong star formation.  This, in 
turn, would affect the colours and morphologies of cluster 
galaxies (e.g., Larson et al.\ 1980; Abadi et al.\ 1999; 
Benson et al.\ 2000; Balogh et al.\ 2000).

Aspects of the stripping of the {\it hot} gaseous halos of 
galaxies in clusters have been considered in previous work (e.g., 
Gisler 1976; Sarazin 1979; Takeda et al.\ 1984).  Mori \& Burkert 
(2000) studied the stripping of dwarf galaxies subject to a 
constant wind using two-dimensional simulations and 
found that the relatively shallow potential wells of these 
systems cannot retain their hot gas component for long.  
However, these authors did not study more massive systems, such 
as normal ellipticals and spirals, where stripping of the 
hot ($\ga 10^6$ K) halo should be much less efficient due to 
their higher masses and deeper potential wells.  [Indeed, a 
recent X-ray survey of massive galaxies in hot clusters by Sun 
et al.\ (2007b) has revealed that {\it most} of the galaxies 
have detectable hot gaseous halos.]  A few other studies have 
examined the stripping of more massive systems but not in the 
context described above.  In particular, they have largely 
focused on the metal enrichment of the ICM 
(e.g., Schindler et al.\ 2005; Kapferer et al.\ 2007), the X-ray 
properties of the galaxies (Toniazzo \& Schindler 2001; Acreman 
et al.\ 2003) or the generation of ``cold fronts'' (e.g., 
Takizawa 2005; Ascasibar \& Markevitch 2006).

In the present paper, we carry out a detailed study of the ram 
pressure stripping of the hot gaseous halos of galaxies as 
they fall into groups and clusters.  This is performed using a 
large suite of controlled hydrodynamic simulations.  Unlike 
most previous studies, we use full three-dimensional (3D) 
simulations in which the galaxies fall into a massive ``live'' 
group or cluster on realistic orbits.  One important aim is to 
derive 
a physically simple and accurate description of the stripping 
seen in the simulations that can be easily employed in the 
modelling of observed or cosmologically-simulated galaxies.  
An additional motivation for deriving such a model is to improve 
the treatment of ram pressure stripping in semi-analytic models 
of galaxy formation.  At present, these models typically 
assume that the hot gaseous halos of galaxies are stripped at 
the instant they cross the virial radius of the group or 
cluster.  Clearly, this is not a realistic assumption, 
especially in the case 
where the mass of the galaxy is not negligible compared to that 
of the group or cluster.  Such semi-analytic models tend to 
predict group and cluster galaxies whose colours are too red 
compared to observations (e.g., Weinmann et al.\ 2006; Baldry et 
al.\ 2006).  If the ram pressure stripping of the hot gaseous 
halos of cluster galaxies is not as (maximally) efficient as 
assumed by these models, the resulting galaxies would be bluer 
and perhaps in better agreement with observations.

The present paper is structured as follows.  In \S 2, we present 
a discussion of our simulation setup and the results of 
convergence 
tests that demonstrate the robustness of our findings.  In \S 3, 
we first outline a simple analytic model for ram pressure 
stripping that is based on the original formulation of Gun \& 
Gott (1972) but which is appropriate for 
spherically-symmetrical gas distributions (as opposed to disks).
We then compare this model to a wide variety of simulations and 
demonstrate that it provides an excellent match to the mass loss 
seen in the simulations.  Finally, in \S 4, we summarise and 
discuss our findings.

\section{Simulations}

To study the ram pressure stripping of galaxies orbiting in 
massive groups and clusters, we make use of the public version 
of the parallel TreeSPH code GADGET-2 (Springel 2005).  By 
default, this code 
implements the entropy-conserving SPH scheme of Springel \& 
Hernquist (2002).  The procedure we use to set up our simulations 
is quite similar to that described in McCarthy et al.\ (2007a) 
(hereafter, M07).  We outline the basic procedure and note any 
relevant differences between our setup and that of M07.

In this study, ram pressure stripping is explored in
two types of simulations.  We refer to the first type as 
the ``uniform medium'' runs, where a galaxy is run through a 
uniform gaseous medium at constant velocity.  In the second type 
of simulations (the ``2-system'' runs), the galaxies 
are placed on realistic orbits through a massive ``live'' galaxy 
group.  In the uniform medium runs, the ram pressure to which
the galaxy is exposed is effectively constant with time.  
Furthermore, there is no external gravitational potential (i.e., 
due to a massive group or cluster) to tidally distort or strip 
the galaxy.  As a result, these simulations should provide a 
pure test of ram pressure stripping and should be easier to 
model than 
the second type of simulations.  On the other hand, if the 
lessons learnt from modelling the uniform medium runs do not 
also generally apply to more realistic situations, such as those 
in the 2-system runs, they will be of little practical use.  
This is why we have elected to use both types of simulations to 
study this problem.

\subsection{Initial conditions and simulation characteristics}

The galaxies (and the groups into which they fall, in the 
case of the 2-system runs) are represented by spherically-symmetric 
systems composed of a realistic mixture of dark matter and 
diffuse baryons.

The dark matter is assumed to follow the NFW distribution 
(Navarro et al.\ 1996; 1997):

\begin{equation}
\rho(r) = \frac{\rho_s}{(r/r_s)(1+r/r_s)^2}
\end{equation}

\noindent where $\rho_s = M_s/(4 \pi r_s^3)$ and

\begin{equation}
M_s=\frac{M_{200}}{\ln(1+r_{200}/r_s)-(r_{200}/r_s)/(1+r_{200}/r_s)} 
\ \ .
\end{equation}

Here, $r_{200}$ is the radius within which the mean
density is 200 times the critical density, $\rho_{\rm crit}$, and
$M_{200} \equiv M(r_{200}) = (4/3) \pi r_{200}^3 \times 200
\rho_{\rm crit}$.

The only `free' parameter of the NFW profile is the scale 
radius, $r_s$.  The scale radius is often expressed in terms of 
a concentration parameter, $c_{200} \equiv r_{200}/r_s$.  By 
default, we adopt the mean mass-concentration ($M_{200}-c_{200}$) 
relation derived from the {\it Millennium Simulation} (Springel 
et al.\ 2005) by Neto et al.\ (2007).  This relationship is 
similar to that derived previously by Eke et al.\ (2001).

For simplicity, the diffuse baryons are assumed to initially 
trace the dark matter distribution, with the ratio of gas to 
total mass set to the universal ratio $f_b = \Omega_b/\Omega_m = 
0.022 h^{-2}/0.3 = 0.141$, where $h$ is the Hubble constant in 
units of $100$ km s$^{-1}$ Mpc$^{-1}$.  The other properties of 
the diffuse gas (i.e., temperature and pressure profiles) are 
fixed by ensuring the gas is initially gravitationally bound and 
in hydrostatic equilibrium,

\begin{equation}
\frac{dP(r)}{dr} = - \frac{G M(r)}{r^2} \rho_{\rm gas}(r) \ \ .
\end{equation}
    
While the assumption that the gas initially traces the dark 
matter is reasonable for the bulk of the baryons in massive 
groups and clusters (e.g., Vikhlinin et al.\ 2006; McCarthy et 
al.\ 2007b), it is almost certainly not a very realistic 
approximation for relatively low-mass systems, such as galaxies.  
The reason, of course, is that non-gravitational physics, such 
as cooling and feedback due, for example, to supernovae and/or 
AGN, which are neglected in our simulations, can significantly 
alter the properties of the gas in these systems.  These 
processes are poorly understood and the properties of the gas 
will likely depend sensitively on the assumed feedback model.  
Therefore, any distribution we select for the hot gaseous halo 
of the galaxies will be somewhat {\it ad hoc}.  The important 
point, however, is that one can use the simulations to 
develop a {\it physical} model for ram pressure stripping that 
can, with some confidence, be applied more generally.  We argue 
that the analytic model developed below is just such a 
model.  As will be demonstrated, tuning the model to match just 
one of our simulations results in very good agreement with all the 
other simulations, in spite of their widely varying physical 
conditions.

\begin{figure}
\centering
\includegraphics[width=8.4cm]{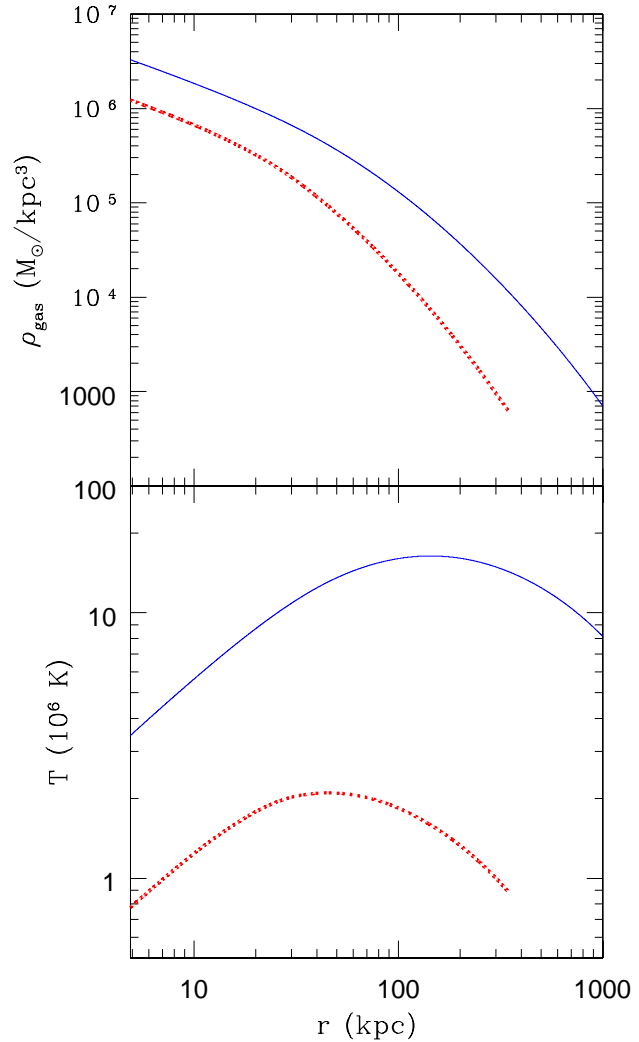}
\caption{The initial gas density (top) and temperature (bottom) 
profiles for the hot halo of a galaxy with mass $M_{200} = 
4\times10^{12} M_\odot$ (dotted red curves) and a group with mass 
$M_{200} = 10^{14} M_\odot$ (solid blue curves).
 }
\end{figure}

The reader is referred to \S 2 of M07 for a detailed discussion 
of how we establish equilibrium configurations of dark matter 
and gas particles that follow an NFW 
distribution\footnote{However, one difference of note is that 
instead of using the dark matter particle positions from our 
isolated runs to set the positions of the gas particles, we now 
morph a glass distribution into the desired NFW profile to set 
the gas particle positions (see \S 2 of M07).  This was done to 
ensure a perfectly `cold' start.  We have run our galaxies and 
groups (with both baryons and dark matter particles in place) in 
isolation for many dynamical times and verified that they are 
stable, unevolving systems.}.  In the case of the 2-system runs, 
the more massive system is set to have a total mass 
of $M_{200} = 10^{14} M_\odot$, while the less massive 
systems have masses in the range $2 \times 10^{12} M_\odot 
\leq M_{200} \leq 10^{13} M_\odot$ (i.e., mass ratios from 50:1 
to 10:1).  Thus, the 2-system runs represent galaxies 
with masses comparable to or larger than a normal elliptical 
falling into a moderate mass group/low mass cluster.  Note that for 
galaxies within this mass range, the mean temperature of their 
gaseous halos ranges between $\approx 1-3\times10^6$ K.  In 
Fig.\ 1, we show the initial gas density and temperature 
profiles for the hot gaseous halo of one of the galaxies and for 
the ICM of the $10^{14} M_\odot$ group.

The default gas particle mass, $m_{\rm gas}$, is set to 
$2\times10^{8} f_b \ M_\odot$, while the default dark matter 
particle mass, $m_{\rm dm}$, is set to $2\times10^{8} 
(1-f_b) \ M_\odot$.  In the 2-system runs, these masses are 
fixed for both the group and the galaxy.  This implies 
that the group is resolved with half a million gas and dark 
matter particles (each) within $r_{200}$.  The gravitational 
softening length for both the gas and dark matter particles is 
set to 5 kpc for all our simulations.  (We have experimented with 
different values of this parameter and find no significant 
differences in the results.)

A standard set of SPH parameters is adopted (e.g., Springel 
2005).  The number of SPH smoothing neighbour particles is set 
to 32, the artificial viscosity $\alpha_{\rm visc}$ parameter is 
set to 0.8 (see \S 2.2), and the Courant timescale coefficient 
is set to 0.1.

The simulation data are output frequently, at intervals of 50 
Myr, 
and the simulations are run for a maximum duration of 10 Gyr in 
the case of the 2-system runs or until a convergent result is 
achieved in the case of the uniform medium runs.

The effects of ram pressure stripping are quantified by
computing the mass of gas that remains gravitationally bound to 
the galaxy as a function of time.  To determine which gas and 
dark matter particles are bound to the galaxy in any particular 
simulation output, we use the iterative method outlined in 
Tormen et al.\ (1998) and Hayashi et al.\ (2003).  Starting from 
the distribution of particles that were bound at the previous 
simulation output (noting that all particles were bound 
initially), the potential, kinetic, and, in the case of the gas, 
the internal energies of each of the particles in the rest-frame 
of the galaxy are computed.  We discard all particles for which 
the sum of the kinetic and internal energies exceeds the 
potential energy.  The rest-frame of the bound structure is 
then recomputed, as are the energies of each particle, and any 
additional unbound particles are identified and discarded. This 
procedure is repeated until no further particles are identified 
as being unbound.  Furthermore, it is implicit that once 
a gas particle has been lost due to stripping it cannot at a 
later time become gravitationally bound again (i.e., the 
mass of bound gas is necessarily a monotonically decreasing 
function of time).  In this way, we are calculating a 
conservative lower limit to the mass of bound gas.

\begin{figure}
\centering
\includegraphics[width=8.4cm]{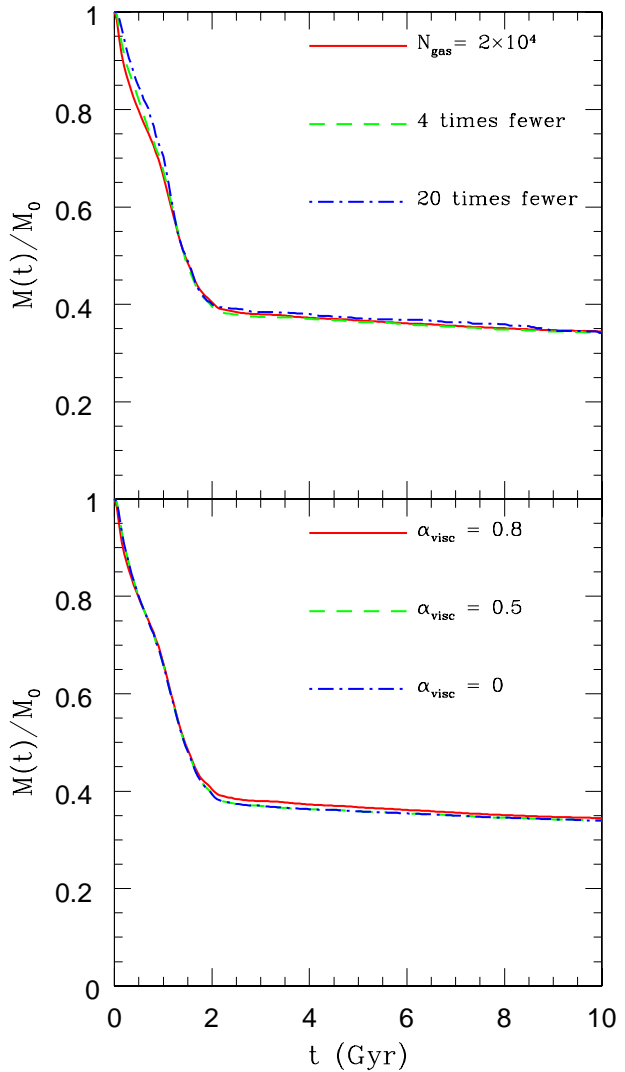}
\caption{The effects of numerical resolution and artificial
viscosity strength on the ram pressure stripping of the galaxy
in
the default 2-system run (see \S 3.3.1).  Plotted is the ratio
of
gravitationally bound gas mass at time $t$ to the initial mass
of gas (at $t = 0$) versus time.  In the top panel, we show the
effect of raising the gas particle mass (i.e., lowering the
particle number) from our default gas mass resolution of $m_{\rm
gas} = 2.82 \times 10^7 M_\odot$.  In the bottom panel, we show
the effect of lowering the SPH artificial viscosity parameter
$\alpha_{\rm visc}$.
 }
\end{figure}

\subsection{Numerical issues}

There are a variety of numerical issues that could potentially
affect the simulations and hamper the development of a physical 
model for ram pressure stripping.  Perhaps of most concern is 
the 
effect of limited numerical resolution and, in the case of SPH 
simulations, the role of the artificial viscosity term, which 
itself is resolution-dependent.  The artificial viscosity, which 
is necessary in order for SPH algorithms to capture shocks, acts 
like an excess pressure for the gas particles in their equation 
of motion and is therefore potentially relevant to our discussion 
of ram pressure stripping.  

We have investigated the effects of numerical resolution and 
artificial viscosity in our default 2-system run (see \S 3.3.1 
for a 
description of this run).  The results are plotted in Figure 2.  
In the top panel, we show the effect of 
degrading the mass resolution of the gas particles (the mass 
resolution of the dark matter particles is the same for all 
these runs) on the ram pressure stripping of the galaxy.
In the default case, there are $2\times10^4$ bound gas particles 
inside $r_{200}$ of the galaxy initially.  Reassuringly, we find 
that lowering the number of particles does not significantly 
affect the resulting bound mass of gas as a function of time.  
This is the case even when the gas halo is represented initially 
by only 1000 particles.  In fact, significant ($> 20\%$) 
differences appear only when the initial gas particle number in 
the galaxy is lowered to a few hundred (not shown).  Note that 
the top panel of Fig.\ 2 implies that our results are not 
strongly sensitive to the artificial viscosity, since this is a 
resolution-dependent quantity.

The bottom panel of Fig.\ 2 adds further confidence that the 
results are robust.  Here, we experiment with lowering the 
$\alpha_{\rm visc}$ parameter, which controls the effective 
`strength' of the artificial viscosity and is proportional 
to the excess pressure assigned to each gas particle in the 
equation of motion.  Lowering 
the value of $\alpha_{\rm visc}$ from the default value of 0.8 
has no significant consequences for the resulting bound mass of 
gas.  This is the case even when the artificial viscosity is set 
to zero\footnote{This may seem somewhat surprising at first 
glance since the galaxy is moving at a high velocity and 
therefore shock heating might be expected to be important (i.e., 
it could raise the entropy of the gas causing some of it to 
become unbound).  However, as discussed in \S 3.1 (see also 
M07), both idealised and cosmological simulations show that shock 
heating of the gas halos of galaxies accreted by groups and 
clusters is unimportant.  Most of the interaction energy is 
thermalised in the ICM of the main halo.}.

We conclude that our ram pressure results are robust to our 
choice of resolution and artificial viscosity strength. It 
should be noted, however, that ram pressure is not 
the only mechanism by which gas can be stripped from galaxies as 
they orbit about groups and clusters.  In particular, 
Kelvin-Helmholtz (KH) and Rayleigh-Taylor (RT) instabilities can 
potentially develop at the interface between the hot halo of the 
galaxy and the ICM and eventually completely disrupt or destroy 
the gaseous halo of the galaxy.  It is known that SPH 
simulations tend to suppress such instabilities in the presence 
of large density gradients across the interface.  This, in 
turn, will make the hot halo of a galaxy more resilient 
to stripping than it otherwise would have been.  
A good example of this can be found in Agertz et al.\ (2007), 
where a comparison between several Eulerian grid-based codes 
(which accurately follow the growth of 
these instabilities) and several Lagrangian SPH codes is 
performed for an idealised case where a `blob' of gas 
moves through a uniform density medium.  For example, their 
Fig.\ 4 shows that, for one particular case, the grid-based 
codes all predict complete disruption of the blob at $t \ga
\tau_{\rm KH}$ (where $\tau_{\rm KH}$ is the KH timescale, i.e., 
the time it takes KH instabilities to fully grow), whereas the 
SPH codes predict that the blob should remain intact.

With this in mind, one might conclude that SPH simulations 
such as ours will overestimate the survivability of the hot halo 
of a galaxy.  However, it is important to note that Agertz 
et al.\ find that the grid-based and SPH-based codes agree with 
each other rather well for $t \la \tau_{\rm KH}$ (see also 
Appendix A of the present study).  Following the 
approach of Mori \& Burkert (2000) (see also Nulsen 1982; Murray 
et al.\ 1993; and Mayer et al.\ 2006), the Kelvin-Helmholtz 
timescale (including the stabilising effects of gravity) can be 
estimated as:

\begin{eqnarray}
\tau_{\rm KH} & = & \frac{F M_0}{\dot{M}_{\rm KH}}\\
   & = & 2.19\times10^9 \ \biggl(\frac{F}{0.1}\biggr)
\biggl(\frac{M_0}{10^9 \ M_\odot}\biggr)^{1/7} \nonumber\\ 
 & & \ \ \times \ \biggl(\frac{n_{\rm 
ICM}}{10^{-4} \ {\rm cm}^{-3}}\biggr)^{-1} \biggl(\frac{v_{\rm
gal}}{10^3 \ {\rm km \ s}^{-1}}\biggr)^{-1} {\rm yr} \ \ , 
\nonumber
\end{eqnarray}

\noindent where $F$ is the baryon fraction of the galaxy, $M_0$ 
is the total mass of the galaxy within the radius down to which 
the galaxy has been stripped by ram pressure, $n_{\rm 
ICM}$ is the number density of hydrogen atoms in the ICM, and 
$v_{\rm orb}$ is the velocity of the galaxy with respect to the 
ICM.

For our default 2-system run (see \S 3.3.1), for example, we 
estimate from eqn.\ (4) that the Kelvin-Helmholtz timescale at 
pericentre is approximately 4.5 Gyr (i.e., which is comparable 
to 
the duration of our simulations).  Since most of the orbital 
period of the galaxy is spent far from pericentre, the value 
of $\tau_{\rm KH}$ will be substantially longer than this.  Note 
also that the timescale associated with the growth of RT 
instabilities is comparable to or exceeds $\tau_{\rm KH}$. 
Therefore, we do not expect KH or RT instability stripping to 
have important consequences for the results or conclusions of 
this study.  We also point out that eqn.\ (4) neglects the 
possibly important effects of radiative cooling, physical 
viscosity, magnetic fields, etc., all of which will tend to 
damp (and possibly halt) the growth of such instabilities in 
real cluster galaxies.

Finally, in order to dispel any lingering doubts that our 
adopted SPH approach is unable to treat ram 
pressure stripping accurately, we have made a direct comparison 
of the predictions of the Lagrangian SPH code GADGET-2 and 
the Eulerian AMR code FLASH for one of our uniform medium runs.  
This comparison is presented in Appendix A and shows that there 
is excellent quantitative agreement between the results of 
the two codes.

\section{Results}

\subsection{Analytic expectations}

The study of ram pressure stripping of galaxies as they fall into 
groups and clusters dates back to the seminal paper of Gunn \& 
Gott (1972).  Using a static force argument, these authors 
derived a simple, physically-motivated condition for the 
instantaneous ram pressure stripping of a gaseous disk moving 
face-on through the ICM.  The gas will be 
stripped if the ram pressure, $P_{\rm ram}$, defined as 
$\rho_{\rm ICM} v_{\rm orb}^2$ (where $\rho_{\rm ICM}$ is the 
density of the ICM and $v_{\rm orb}$ is the speed of the galaxy 
with respect to the ICM), exceeds the gravitational restoring 
force per unit area on the disk, which they derive as $2 \pi 
G \Sigma_* \Sigma_{\rm gas}$ (where $\Sigma_*$ and $\Sigma_{\rm 
gas}$ are the stellar and gaseous surface densities of the disk, 
respectively).  We now seek to derive an analogous model for 
the ram pressure stripping of a spherically-symmetric gas 
distribution.  

\begin{figure}
\centering
\includegraphics[width=8.4cm]{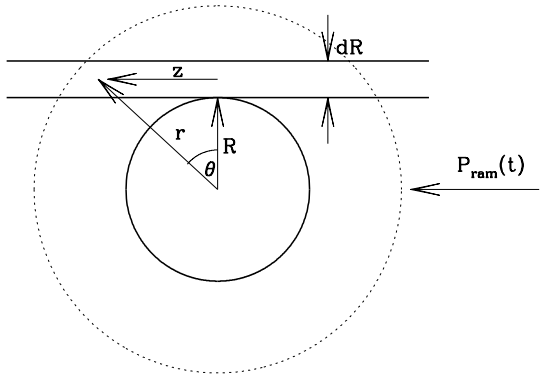}
\caption{A schematic diagram of the ram pressure stripping of a
spherically symmetric gas distribution.  Here, the ram pressure
force is directed from left to right and we consider the ratio of
the ram pressure force to the gravitational restoring
force per unit area for a projected annulus of width $dR$
at the outer edge (radius $R$) of the gaseous halo of the galaxy.
 }
\end{figure}

Since it is the least bound material, gas at the outer 
projected edge of the system will be stripped first (see the 
schematic diagram in Figure 3).  Consider gas in a projected 
annulus between radii $R$ and $R+dR$.  The projected 
area of this annulus, $dA$, is $2 \pi R dR$.  Therefore, the 
force due to ram pressure on this annulus is simply $F_{\rm ram} 
= P_{\rm ram} dA$.  The annulus of gas will be displaced in the 
direction opposite to $v_{\rm orb}$ (which we will call the $z$ 
direction) and will be stripped if the force due to the ram 
pressure exceeds the maximum gravitational restoring force in 
this direction.  The maximum gravitational restoring 
force, $F_{\rm grav}$, can be written approximately as $g_{\rm 
max}(R) \Sigma_{\rm gas}(R) dA$, where $g_{\rm max}(R)$ is the 
maximum 
restoring acceleration in the $z$ direction and $\Sigma_{\rm 
gas}(R)$ is the projected surface density of the gas in the 
annulus.  Therefore, the ram pressure stripping condition can be 
written as:

\begin{equation}
\rho_{\rm ICM} v_{\rm orb}^2 > g_{\rm max}(R) \Sigma_{\rm gas}(R) 
\ \ .
\end{equation}

If the gas density and total mass profiles of the galaxy can be 
represented by simple power laws, it is straightforward to 
evaluate the right-hand side of equation (5).  In the case of a 
singular isothermal sphere, for example, where $\rho_{\rm 
gas}(r) \propto r^{-2}$ and $M_{\rm gal}(r) \propto r$ 
(where $M_{\rm gal}(r)$ is the {\it total} mass within radius $r$), 
we find $g_{\rm max}(R) = G M_{\rm gal}(R) / (2 R^2)$ and 
$\Sigma_{\rm gas}(R) = \pi \rho_{\rm gas}(R) R$.  This leads to 
the following stripping condition 

\begin{equation}
P_{\rm ram}(t) > \frac{\pi}{2} \frac{G M_{\rm gal}(R) \rho_{\rm 
gas}(R)}{R} \ \ .
\end{equation}

For more general gas density and total mass profiles, the 
condition for ram pressure stripping may be expressed as

\begin{equation}
P_{\rm ram}(t) > \alpha \frac{G M_{\rm gal}(R) \rho_{\rm 
gas}(R)}{R} \ \ ,
\end{equation}

\noindent where $\alpha$ is a geometric constant of order unity 
which depends on the precise shape of the gas density 
and total mass profiles of the galaxy.  We note that equation (7) 
is similar to the analytic stripping conditions derived 
previously by Gisler (1976) and Sarazin (1979) (among others) 
for elliptical galaxies.  

Equation (7) implies that all the gas beyond the 3D radius 
$R_{\rm strip}$ where the ram pressure exceeds the gravitational
restoring force per unit area (which we will refer to as the 
stripping radius) will be stripped.  By assumption, 
the properties of both the gas and dark matter within the 
stripping radius are unmodified by the stripping.  The 
left-hand side of eqn.\ (7) makes it clear that the ram 
pressure is, in general, a function of time (i.e., for 
non-circular orbits).

Below, we use the idealised uniform medium runs to test 
this simple analytic model.  However, before doing so it is 
worth briefly discussing some of the assumptions of this simple 
model and their validity.  Firstly, the model neglects KH and RT 
instability stripping but, as we argued in \S 2.2, we do not 
expect this to be an important omission.  Perhaps of more 
concern is that, by assuming that the properties of the system 
within the stripping radius do not change with time, the model 
implicitly neglects environmental effects such as tidal 
stripping and gravitational shock heating.  In Appendix B, we 
show, using a simple argument, that one expects ram pressure 
stripping to be more efficient than tidal stripping for cases 
where the mass of the galaxy is less than about 10\% of the mass 
of the group.  In other words, for galaxies with masses of less 
than about 10\% of the group mass, tidal stripping is not 
expected to substantially modify the structure of the galaxy 
within its stripping radius.  Our 2-system runs involve only 
systems with mass ratios $\ge$ 10:1.

The neglect of shock heating would naively appear to be a more 
serious omission, since the commonly-held picture of structure 
formation is that gas accreted by a massive system is 
shocked at the virial radius up to the virial temperature of the 
massive system.  Thus, one might expect the hot gas halo of the 
galaxy to be quickly shock heated and become unbound.  However, 
high resolution simulations (both cosmological and idealised) 
do not confirm this picture.  In particular, if the material 
being accreted is in small dense ``lumps'' (e.g., low-mass 
virialised systems, as in the present case), it can penetrate all 
the way to the core of the massive system without being 
significantly shocked (e.g., Motl et al.\ 2004; Murray \& Lin 
2004; Poole et al.\ 2006; M07; Dekel \& Birnboim 2007).  In 
fact, most of the interaction energy is thermalised in the 
ambient medium of the more massive system (the ICM, in this 
case), while the accreted gas sinks to bottom of the 
potential well (see M07 for a detailed discussion). However, 
M07 found that the fraction of the total energy that is 
thermalised in the gas of the less massive system (the 
galaxy, in this case) increases almost linearly with the ratio 
of the mass of the less massive system to the total mass of both 
systems.  Therefore, shock heating {\it is} expected to become 
important for cases where the mass of the galaxy is comparable to 
the mass of the group.  Our 2-system runs, however, only involve 
galaxies with masses lower than 10\% of the mass of the group.

\subsection{Uniform medium runs}

We now explore the ram pressure stripping of galaxies as they 
move through a uniform density gaseous medium.  For the uniform 
medium, we select densities that are typical of the group/cluster 
environment.  The temperature of the medium is set such that its 
pressure equals that of the hot halo of the galaxy at its outer 
edge (i.e., the gaseous halo would be static if it were 
not moving with respect to the uniform medium).  The galaxies 
are assigned velocities typical of systems orbiting in genuine 
groups and clusters (i.e., comparable to the circular velocity 
of the group or cluster).

In Fig.\ 4 we plot the bound mass of gas as a function of time 
for a small selection of the uniform medium runs we have 
performed and compare this with our proposed analytic model.  We 
focus first on the $M(t)$ curves from the simulations 
(solid red curves).  Firstly, the $M(t)$ curves in both panels 
asymptote to a particular value, as one would expect from the 
physical model proposed above where the ram pressure is 
effectively held constant with time but is low enough that 
not all of the gas should be stripped.  In the bottom panel, 
where a galaxy is moved through media of two different densities 
but with velocities chosen such that the ram pressure is the 
same, the resulting $M(t)$ curves are very similar.  This 
unambiguously demonstrates that the mass loss is 
indeed due to ram pressure stripping.  

\begin{figure}
\centering
\includegraphics[width=8.4cm]{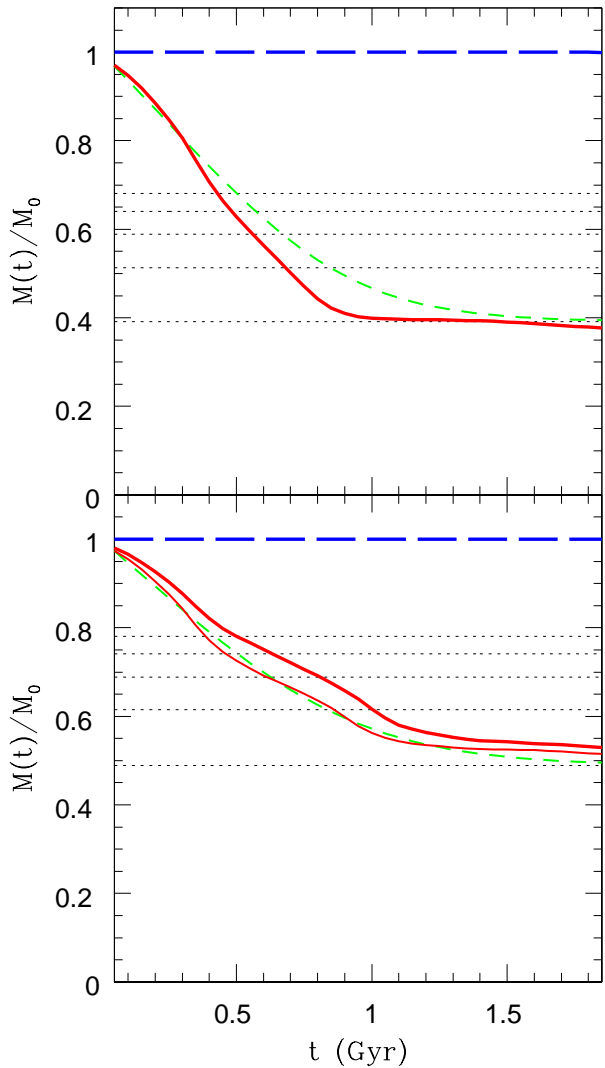}
\caption{An example of ram pressure stripping in the uniform
medium simulations.  In the top panel, a galaxy of mass
$M_{200}=4\times10^{12} M_\odot$ is run through a uniform
gaseous medium of density $100 f_b \rho_{\rm crit}$ at a
velocity of 1000 km s$^{-1}$.  The solid red and dashed blue
curves show the bound mass of gas and dark matter (respectively)
in the simulation.  In the bottom panel, the same galaxy is
run through two different media: the thick red curve
corresponds to the case where the background density is the same
as in the top panel, but the velocity is 760 km s$^{-1}$, while
the thin red curve corresponds to the case where the velocity
is 1000 km s$^{-1}$ but the density is a factor of
$(1000/760)^2$ times lower than in the top panel.  Thus, the ram
pressure is the same for both cases in the bottom panel.  In
both the top and bottom panels the horizontal dotted lines
correspond to the predictions of equation (7) for $\alpha =$ 2,
4, 6, 8, and 10 (bottom to top).  The green dashed curve
corresponds to equation (7) with $\alpha = 2$ but with a time
delay factor that accounts for how long it takes the for galaxy
to respond to ram pressure stripping (i.e., approximately a
sound crossing time), as discussed in the text.
 }
\end{figure}

A comparison to the predictions of equation (7) (horizontal 
dotted lines) demonstrates that the asymptotic behaviour 
of the simulations is reproduced if  $\alpha \approx 2$.  (Note 
that this is very similar to the analytic estimate of $\pi/2$ 
derived in \S 3.1 for an isothermal sphere.)  In fact, all the 
uniform medium runs we have performed yield a value of $\alpha$ 
close to 2.  However, it is immediately apparent that the 
approximation of instantaneous stripping is not a particularly 
good one.  For example, in the cases plotted in Fig.\ 4 it 
takes $\sim 1$ Gyr of stripping to reach a convergent value 
(i.e., to reach the 3D radius where the ram pressure equals the 
gravitational force per unit area).  This ``time delay'' 
has been noted previously in studies of the stripping 
of cold disks (e.g., Roediger \& Br{\"u}ggen 2006; 2007) and is 
expected on physical grounds; the hot halo of the galaxy can 
only respond to changes in the local environment on a finite 
timescale.  What is the relevant timescale?  If the galaxy is 
moving subsonically, a natural choice might be the sound 
crossing time; i.e., the time it takes for a pressure wave to 
cross the galaxy's hot halo.  If the galaxy is moving 
supersonically, a better choice might be the time it takes a 
forward shock to propagate across the galaxy (e.g., Nittmann et 
al.\ 1982; Mori \& Burkert 2000).  (Although, as we noted 
above, shock heating of the hot gas of the galaxy is minimal in 
our simulations.)  Alternatively, Roediger \& Br{\"u}ggen 
(2007) estimate and use the timescale required for the ram 
pressure to accelerate the gas to the galaxy's escape velocity.  
We have experimented with including a time delay factor into 
the analytic model (how we do this is described below) that is 
set to either of these three timescales times.  In practice, we 
find that use of either of these timescales leads to very 
similar results.  This is not too surprising.  The similarity 
between the sound and shock crossing times is due to the fact 
that, in the rest frame of the group, the galaxy is typically 
orbiting at transonic velocities (i.e., Mach number $\sim 1$).  
The similarity between the sound crossing time and the time 
required to accelerate gas to the galaxy's escape velocity is 
also not coincidental.  Since the galaxy's hot halo is in 
approximate hydrostatic equilibrium, the mean temperature 
of the gas will be close to the overall virial temperature of 
the galaxy (which is dominated by the mass in dark matter) and 
therefore the sound crossing time of the hot halo will be of 
order the dynamical time of the galaxy.  Consequently, if the 
force due to the ram pressure is of order the gravitational 
restoring force, as is the case for typical transonic 
velocities, the time it takes to accelerate the gas to the 
escape velocity will be comparable to the sound crossing time.  
Note, however, that if the galaxy's motion is highly supersonic 
(or if the gas is not in equilibrium) one might expect 
differences between the three timescales.  The present study 
does not consider this regime and instead focuses 
on the more physically relevant transonic regime where all 
three timescales are similar.  Below, we present results 
based on using the sound crossing time only.

Note that the simulation $M(t)$ curves plotted in Fig.\ 4 show 
that the mass loss proceeds with time more or less linearly until 
convergence is achieved.  (This is generally true 
of the 2-system runs presented below, as well.)  We therefore 
assume that the mass of gas stripped over some time 
interval $\Delta t$ is just the total mass of stripped gas 
inferred from the instantaneous assumption (i.e., the total gas 
mass external to the stripping radius) scaled by the ratio 
$\Delta t / t_{\rm ram}$, where $t_{\rm ram}$ is the 
characteristic timescale for ram pressure stripping (i.e., 
approximately the sound crossing time).  For an appropriate 
comparison to the simulations, we set $\Delta t$ to the adopted 
simulation output time interval of 50 Myr.

The sound crossing time of the gaseous halo at any particular 
time is calculated as:

\begin{equation}
t_{\rm sound} = \int_0^R \frac{dr'}{c_s(r')}
\end{equation}

\noindent where $R$ is the maximum radial extent of the 
bound galactic gas and  $c_s(r)$ is the local sound speed 
profile, which is given by $[\gamma P_{\rm gas}(r)/\rho_{\rm 
gas}(r)]^{1/2}$ with $\gamma = 5/3$.  Note that for an 
isothermal gas this leads to the familiar relation $t_{\rm sound} 
= R/c_s$.

In fact, the time it takes the gaseous halo to respond to 
changes in the local environment will only be comparable to 
the sound crossing time, not exactly equal to it.  We therefore 
multiply 
this timescale by an adjustable coefficient $\beta$ (which 
will be of order unity) when computing how much mass can be 
stripped over a time interval (i.e., $t_{\rm ram} = \beta t_{\rm 
sound}$).

\begin{figure}
\centering
\includegraphics[width=8.4cm]{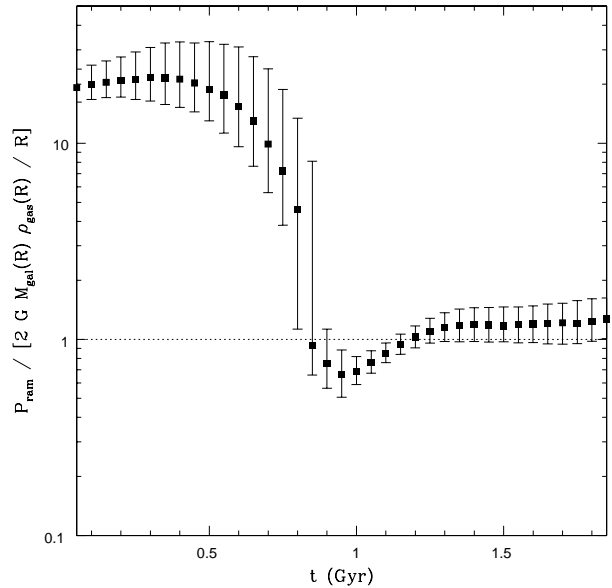}
\caption{The ratio of ram pressure to gravitational restoring
force per unit area (assuming $\alpha = 2$) at the outer edge of
the gaseous halo of the galaxy plotted in the top panel of Fig.\
4, as a function of time.  The solid squares represent the 
median
of the 500 outermost gravitationally bound gas particles while
the error bars represent the 25th and 75th percentiles.  After 
$t\approx 0.8$ the ram pressure and restoring force per unit 
area
become comparable, which is why mass loss ceases after this time
in Fig.\ 4.
 }
\end{figure}

The resulting model is plotted in Fig.\ 4.  In this case 
$\alpha$ 
has been fixed to $2$ to obtain agreement with the asymptotic 
$M(t)$ behaviour of the simulated galaxies.  A value of $0.5 < 
\beta < 0.7$ yields good agreement with the rate of decline of 
the bound gas mass seen early on in the simulations (shown is 
the case corresponding to $\beta = 2/3$).  It is worth bearing in 
mind that the analytic model uses only the {\it initial} radial 
profiles of the galaxy to compute the bound mass of gas as a 
function of time.  The fact that the model matches the 
simulations and that the required values of $\alpha$ and $\beta$ 
are of order unity is encouraging.

As a further test of the analytic model, in Fig.\ 5 we plot the 
ratio of ram pressure to the restoring force per unit area 
(assuming $\alpha = 2$) at the outer edge of the gaseous halo of 
the simulated galaxy examined in the top panel of Fig.\ 4, as a 
function of time.  This plot clearly demonstrates that at early 
times the ram pressure exceeds the gravitational restoring force 
per unit area, which is why stripping occurs.  As shown in the 
top panel of Fig.\ 4, stripping continues until $t \approx 0.8$ 
Gyr and then stops rather abruptly.  With Fig.\ 5 we now clearly 
see the reason for this behaviour: the ram pressure 
no longer exceeds the restoring force per unit area at the outer 
edge of the bound halo after this time.  In addition, we confirm 
that the maximum radial extent of the bound gas at $t > 0.8$ Gyr 
corresponds closely to the 3D radius where ram pressure equals 
the restoring force per unit area calculated from the {\it 
initial} gas distribution.  This validates the basic assumptions 
of our analytic model, outlined in \S 3.1.

Having calibrated the analytic model against the uniform medium 
simulations (i.e., $\alpha$ and $\beta$ are now fixed), we now 
proceed to see whether or not this simple physical model can 
also account for the mass loss in the more realistic 2-system 
runs.

\subsection{2-system runs}

\subsubsection{The default 2-system run}

\begin{figure}
\centering
\includegraphics[width=8.4cm]{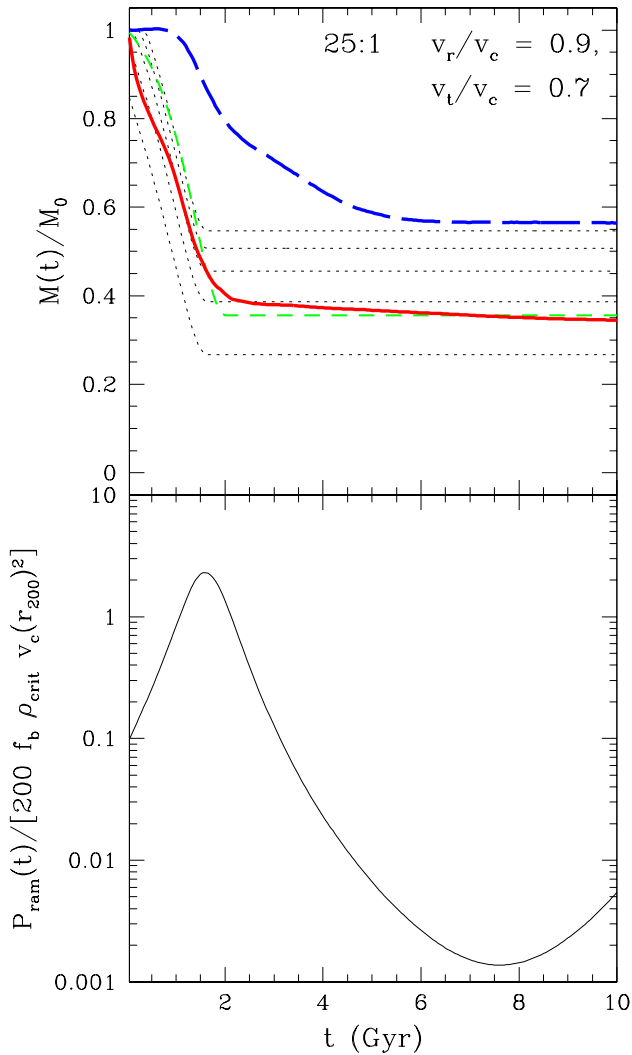}
\caption{Ram pressure stripping in the default 2-system run.
Top
panel: the solid red and dashed blue curves show the bound mass
of gas and dark matter, respectively, in the simulation.  The
green dashed curve corresponds to predictions of the analytic
model (for $\alpha = 2$ and $\beta = 2/3$) where stripping
occurs
on approximately a sound crossing time.  The dotted curves are
the predictions of equation (7) with $\alpha =$ 2, 4, 6, 8, and
10 (bottom to top) under the assumption of instantaneous
stripping.  Bottom panel: the ram pressure as a function of time
as the galaxy orbits the group.  The ram pressure has been
normalised to the characteristic value of $\overline{\rho_{\rm
ICM}} v_c(r_{200})^2$.  For this particular orbit, which
corresponds to the most common orbit of infalling substructure
in cosmological simulations, pericentric (apocentric) passage
occurs at $t \approx 1.5$ Gyr ($t \approx 7.5$ Gyr).
 }
\end{figure}

The default 2-system run follows a massive galaxy with $M_{200} 
= 4\times10^{12} M_\odot$ falling into a moderate mass group of 
$M_{200} = 10^{14} M_\odot$ (implying a mass ratio of 25:1).  As 
noted in \S 2, the concentration of these systems is set to 
match the mean mass-concentration of dark matter halos in the 
{\it Millennium Simulation}.  The 2-system runs are initialised 
such that the virial radii (here defined as $r_{200}$) of the 
two systems are just barely touching.  The adopted orbital 
parameters of the default run correspond to the most common 
orbit of infalling substructure measured in a large suite of 
cosmological simulations by Benson (2005; see his Fig.\ 2).  
Specifically, the initial relative radial velocity component, 
$v_r$, is set to $0.9 v_c(r_{200})$ and the initial relative 
tangential component, 
$v_t$ is 
set to $0.7 v_c(r_{200})$, where $v_c(r_{200})$ is the circular 
velocity of the group at $r_{200}$.  This 
corresponds to a total relative velocity of $\approx 1.1 
v_c(r_{200})$, which agrees well with the results of several 
other similar numerical studies (e.g., Tormen 1997; Vitvitska et 
al.\ 
2002; Wang et al.\ 2005).  In the following subsections, we 
experiment with varying the orbit, mass, and internal structure 
of the galaxy to test the generality of the analytic model.

\begin{figure}
\centering
\includegraphics[width=8.4cm]{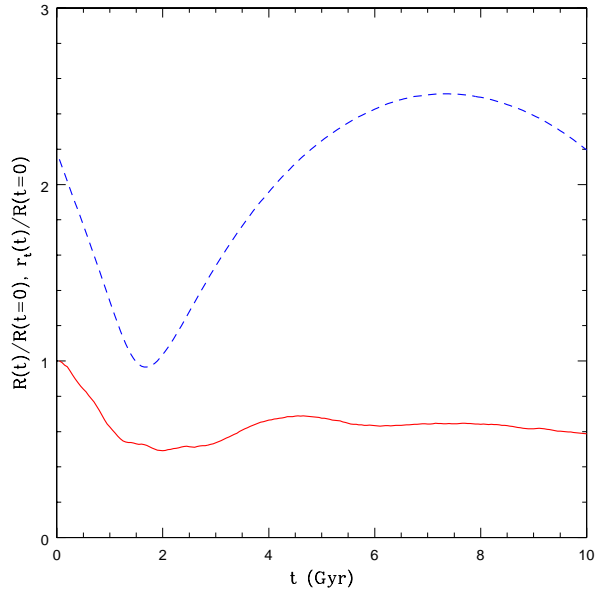}
\caption{The evolution of the galaxy's tidal radius (dashed blue
curve) and the radial extent of its bound gaseous halo (solid
red curve) for the default 2-system run.  The radial extent of
the gas is defined here as the radius enclosing 90\% of
galaxy's bound hot halo.  The tidal radius is larger than the
bound hot gaseous halo by at least a factor of two at all times.
 }
\end{figure}

As in the uniform medium runs, the analytic model is supplied 
with the initial conditions of the galaxy (i.e., its gas and 
dark matter radial profiles) and the magnitude of the ram 
pressure.  In contrast to the uniform medium runs, however, the 
ram pressure is not constant with time.  Using the orbit from 
the simulations, along with the density profile of the group, 
$P_{\rm ram}(t)$ is calculated and passed to the analytic model.  
The analytic model can then predict $M(t)$ once the values of  
$\alpha$ and $\beta$ have been selected.

The mass loss curves for the default 2-system run are plotted
in the top panel of Fig.\ 6.  Overall, the simple analytic model 
with $\alpha = 2$, $0.5 < \beta < 0.7$ (shown is $\beta = 
2/3$) and $t_{\rm ram} = \beta 
t_{\rm sound}$ reproduces the mass loss seen in the default 
2-system run very well.  For example, both the simulations and 
the model show evidence for near convergence in $M(t)$ at $t 
\ga 1.5$ Gyr, which corresponds to the (first) pericentric 
passage and, therefore, to the maximum ram pressure which the 
galaxy experiences along its orbit (see the bottom panel of Fig.\  
6).

\begin{figure*}
\centering
\includegraphics[width=18.5cm]{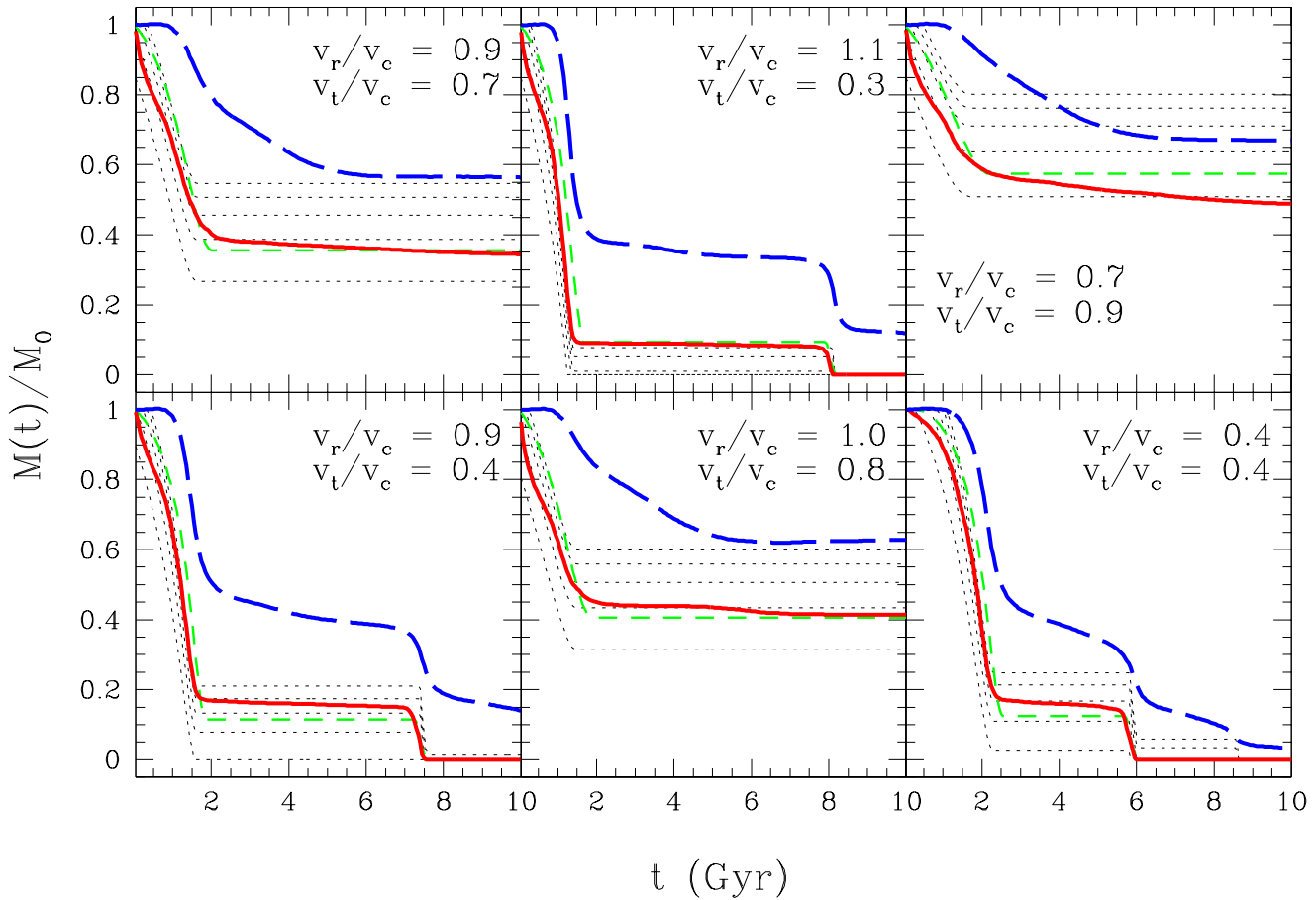}
\caption{Ram pressure stripping as a function of initial
orbital
parameters.  Shown are the mass loss curves of a galaxy with
an initial mass $M_{200} = 4\times10^{12} M_\odot$ falling
into a group with mass $M_{200} = 10^{14} M_\odot$.  Each
panel represents a different orbit, as described in the text.
The line types have the same meanings as in Fig.\ 6.}
\end{figure*}

The analytic model slightly underestimates the mass loss seen in 
the simulations at early times.  This is a result of the fact 
that the hot halo of the galaxy is initially slightly 
over-pressurised with respect to the surrounding hot halo of the 
group. (Note that this was not the case for the uniform medium 
simulations plotted in Fig.\ 4.)  This leads to some expansion 
of the outer gas which, in turn, makes it more susceptible to 
stripping.  Since this effect is in general small and is an 
artifact of our idealised setup, we do not attempt to model it.

While the analytic model with a time delay factor matches the 
simulations well, an instantaneous stripping model with $\alpha 
\approx 4$ (represented by the second dotted curve from the 
bottom) also performs well.  However, even if the agreement 
is reasonable, this model is without physical justification 
and should not be expected to apply in situations that differ 
significantly from those of the default run.  Indeed, this is 
indicated by the results presented later in the paper (c.f. 
Fig.\ 8).

We also note that a significant fraction of the dark 
matter halo is also stripped, particularly near the first 
pericentric 
passage.  This is not unexpected and is due to the tidal forces 
acting on the dark matter.  We do not attempt to model the 
stripping of the dark matter, as there are already several 
published analytic studies which reproduce the dark matter 
stripping and tidal heating in simulations well (e.g., Taylor \& 
Babul 2001; Benson et al.\ 2002).  Instead, the analytic model 
proposed in \S 3.1 simply assumes that, within the stripping 
radius, the properties of the galaxy are unchanged from their 
initial state.  Thus, the dark matter halo is assumed to 
maintain its initial NFW configuration within this radius.
In Appendix B, we present a simple analytic argument that 
validates this assumption for systems where the mass of the 
galaxy is less than about 10\% of the mass of the group.   
We have also directly computed the evolution of the tidal radius 
($r_t$, defined in Binney \& Tremaine 1987; see also Appendix B) 
of the galaxy in the simulations as a function of time.  In 
Fig.\ 7  we compare the tidal radius with the radial extent of 
the hot gaseous halo.  The tidal radius shrinks at pericentre 
and then expands but at all times is safely larger than the 
gaseous halo by at least a factor of 2.

\subsubsection{Varying the orbit of the galaxy}

In Fig.\ 6 we examined the ram pressure stripping of a galaxy 
on the most common orbit seen in cosmological simulations.  We 
now experiment with varying the initial orbital parameters.  This 
will have the effect of changing both the shape and normalisation 
of $P_{\rm ram}(t)$. We use Fig.\ 2 of Benson (2005) to select a 
range of cosmologically likely orbits; the initial velocity 
of some orbits is dominated by the radial component while 
others have nearly circular motions initially\footnote{In fact, 
unlike the other cases, the orbit with $v_r/v_c(r_{200}) = 
v_t/v_c(r_{200}) = 0.4$ is not a common one.  We have simulated 
this atypical orbit just to see if the model breaks down for 
extreme cases.}.  We plot the mass loss curves for six such 
orbits in Fig.\ 8.

The mass loss curves in Fig.\ 8 exhibit a variety of behaviours.  
Orbits that initially have a significant tangential component 
(and have a total velocity of $\sim v_c$) typically undergo 
only one pericentric passage over the course of 10 Gyr.  
Consequently, their associated $M(t)$ curves tend to show 
only one period of significant decline.  Orbits that are 
predominantly radial, on the other hand, typically undergo two 
or more pericentric passages, with each successive passage 
bringing the galaxy closer to the centre of the group.  In these 
cases we see two (or more) periods of significant decline in 
the bound mass of gas, as expected.

In spite of the widely varying orbits, the simple analytic model 
with $\alpha = 2$, $0.5 < \beta < 0.7$ (shown is $\beta =
2/3$), and $t_{\rm ram} = \beta  t_{\rm sound}$ performs 
remarkably well in predicting the mass loss seen in the 
simulations.  For all orbits and at all times the model predicts 
$M(t)$ to within $\approx 10\%$ accuracy. 

Finally, it is interesting to note that if the standard (but 
unphysical) instantaneous ram pressure stripping model were 
adopted, the implication would be that $\alpha$ should vary as 
a function of the orbit.  In particular, from Fig.\ 8, one 
would infer relatively low values of $\alpha$ ($\sim 2-5$) for 
more circular orbits and relatively high values of $\alpha$ 
($\sim 6-10$) for more radial orbits. However, $\alpha$ is 
a geometric constant that is not expected to depend on the 
orbit.  This consideration provided one of the original 
motivations for us to 
explore models where the stripping is not instantaneous.  As we 
have demonstrated, a fixed value of $\alpha \approx 2$ works 
well for all orbits when one takes into account the finite 
time required for stripping.

\subsubsection{Varying the mass of the galaxy}

We now investigate variations in total mass of the galaxy.  This 
will 
mainly have the effect of changing the gravitational 
restoring force (per unit area) of the galaxy at all radii by a 
constant 
factor.  As indicated by Fig.\ 9, the analytic model matches the 
higher mass ratio interactions (lower galaxy masses) well at all 
times but does less well for the low mass ratio 10:1 at late 
times.  In particular, the analytic model predicts that there 
ought to 
be no further stripping following first pericentric passage 
while the simulations show evidence for further stripping.  What 
is the origin of this behaviour?

\begin{figure}
\centering
\includegraphics[width=8.4cm]{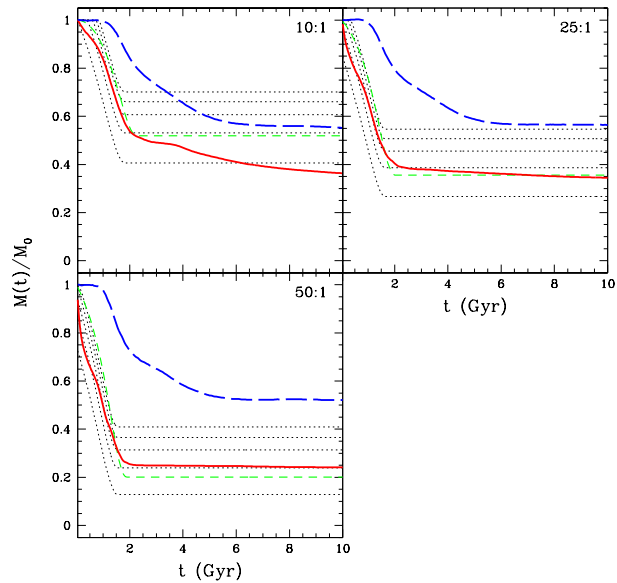}
\caption{Ram pressure stripping as a function of galaxy mass.
Shown are the mass loss curves for a galaxy of varying mass
but with the same initial orbital parameters as in the default
2-system run.  Each panel corresponds to galaxies with
different
total masses as is indicated by the mass ratio in the legend.
(Note that the group mass is fixed at $M_{200} = 10^{14}
M_\odot$
and the default case corresponds to a mass ratio of 25:1.)  The
line types have the same meanings as in Fig.\ 6.
 }
\end{figure}

Inspection of the 10:1 simulation reveals that the 
gaseous halo of the galaxy undergoes significant expansion at 
late times while the analytic model uses the initial gas 
distribution (see Fig.\ 10).  The expansion, in turn, 
makes the 
gas more 
susceptible to ram pressure stripping, and this accounts for the 
decline in the bound gas mass at late times.  The physical 
reason for the expansion of the gaseous halo is as follows.  At 
early times, the ram pressure exceeds the restoring force per 
unit area of the outer gas, which leads to stripping.  This 
stripping proceeds until pericentric passage, when $P_{\rm ram}$ 
is largest.  The remaining bound gaseous 
halo following pericentric passage is of higher mean density and 
pressure than the initial system, since all of the low density 
(less bound) material has been removed.  Following pericentre, 
the galaxy moves out to large group-centric radii, where the 
pressure and density of the ICM are relatively low compared to 
pericentre.  As a result, the gaseous halo of the galaxy 
becomes over-pressurised with respect to the local ICM and begins 
to expand.  This effect is larger in the 
case of more massive galaxies since they are more 
over-pressurised with respect to the ICM.  The expansion 
proceeds until approximately apocentre is 
reached, at which point the galaxy begins to move back into 
denser and higher pressure regions of the group.  (This effect 
is also responsible for the mild decline in bound gas mass for 
the highly tangential orbital case plotted in the top right-hand 
panel of Fig.\ 8.)  It is important to note that 
this over-pressurisation effect is not a numerical artifact, it 
is a real effect that should be experienced by massive galaxies 
with orbits that have large energies and tangential components.

\begin{figure}
\centering
\includegraphics[width=8.4cm]{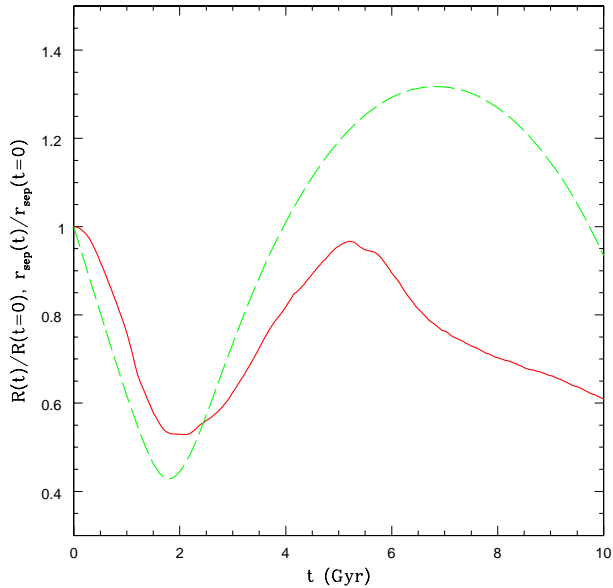}
\caption{The evolution of the radial extent of bound gaseous 
halo (solid red curve) for the 10:1 2-system run plotted in 
Fig.\ 9.  Also shown is the distance ($r_{\rm sep}$) between the 
centres of the 
galaxy and group as a function of time.  Following pericentric 
passage, the hot halo of the galaxy is over-pressurized compared to 
the ambient ICM and begins to expand.  This 
expansion leads to further ram pressure stripping at late times.
}
\end{figure}

Modelling this effect may be possible with some effort.  The 
expansion of the gaseous halo at late times 
is adiabatic, which greatly simplifies matters.  One could 
therefore compute the radial properties of the gaseous halo as a 
function of time using the Lagrangian entropy distribution of 
the gas and assuming hydrostatic equilibrium with an outer 
boundary condition that the pressure must match that of the 
ambient ICM.  However, this procedure is complicated by the fact 
that one must also know the distribution of the galaxy's dark 
matter halo out to the radius of maximum expansion.  While the 
dark 
matter profile at small and intermediate radii is sufficiently 
similar to the initial distribution, this is not the case at 
very large radii.  A proper treatment therefore requires that we 
factor in dark matter stripping and heating.  This could 
potentially be achieved by combining our analytic ram 
pressure model with existing analytic models of dark matter 
stripping and heating (e.g., Taylor \& Babul 2001; Benson et al.\ 
2002).  However, this is beyond the scope of the present study 
and we leave it for future work.  

Finally, we stress that the expansion effect just described is
relevant to cases where both of the following are true: (1) the 
mass of the galaxy is greater than about 10\% of the mass of the 
group; and (2) the orbit has an appreciable tangential component 
and a large enough energy such that apocentre occurs at large 
group radius\footnote{For radial orbits, on the other hand, 
apocentre lies at smaller group radius and, as a result, the 
galaxy does not become over-pressurised with respect to the 
ICM.  In these cases, the 
analytic model matches the mass loss in the simulations quite 
well.} (i.e., comparable to the 
group virial radius).  However, we expect that both of these 
conditions are rarely fulfilled simultaneously in real systems, 
as massive satellites tend preferentially to fall into groups 
and clusters on nearly radial orbits (i.e., along filaments; 
see, e.g., Benson 2005).

\begin{figure}
\centering
\includegraphics[width=8.4cm]{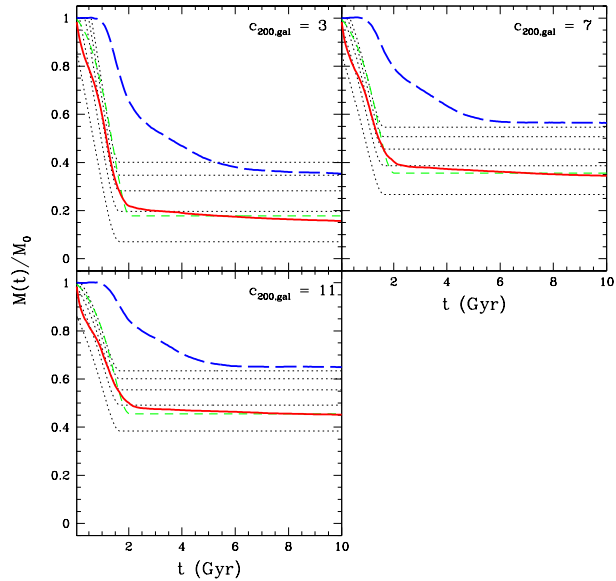}
\caption{Ram pressure stripping as a function of galaxy
internal
structure.  Shown are the mass loss curves of a galaxy with
the same mass and orbit as the default run but with a varying
concentration.  Each panel corresponds to a different
concentration parameter for the galaxy, with the default case
corresponding to $c_{200} = 7$.  The line types have the
same meanings as in Fig.\ 6.
 }
\end{figure}

\subsubsection{Varying the concentration of the galaxy}

Finally, we experiment with varying the internal structure 
of the galaxy (both its gas and dark matter) by varying its 
initial concentration parameter, $c_{200}$ (equivalently, its 
scale radius, $r_s$).  This will mainly have the effect of 
changing the shape of the radial profile of the restoring force 
(per unit area).  This test is 
motivated by the 
fact that in cosmological simulations there is a large degree of 
intrinsic scatter in the concentration parameter for a system of 
fixed mass (e.g., Dolag et al.\ 2004; Neto et al.\ 2007).  Note 
that changing the concentration can also mimic the addition of 
another mass 
component to the galaxy, such as a stellar component (which we 
have neglected to include explicitly).

Fig.\ 11 shows that the concentration has a significant effect 
on the amount of gas that the galaxy is able to retain as it 
orbits about the group.  As expected, as the concentration is 
increased so too is the bound mass of gas.  As in the previous 
experiments, the simple analytic model with $\alpha = 2$, $0.5 
< \beta < 0.7$ (shown is $\beta = 2/3$), and $t_{\rm ram} = 
\beta t_{\rm sound}$ matches the mass loss in the simulations 
very well.

\section{Summary and Discussion}

Using a suite of carefully controlled 3D hydrodynamic 
simulations, we have investigated the ram pressure stripping of 
hot gas in the halos of 
galaxies as they fall into groups and clusters.  We have proposed 
a 
physically simple analytic model that describes the stripping 
seen in the simulations remarkably well.  This model is 
analogous to the original formulation of Gunn \& Gott (1972), 
except that it is appropriate for the case of a spherical gas 
distribution (as opposed to a face-on disk) and takes into 
account that stripping is not instantaneous but occurs on 
approximately a sound crossing time.  The only pieces of 
information that the model requires are the initial conditions of 
the orbiting galaxy (its gas and dark matter profiles), 
the density profile of the ICM and the orbit [the latter two are 
needed to calculate $P_{\rm ram}(t)$].  The model contains two 
tunable 
coefficients that are of order unity.  Fixing these coefficients 
to match the stripping in just one of our idealised 
uniform medium simulations (see \S 3.2) leads to excellent 
agreement with all our other simulations.  With 
the exception of cases where the mass of the galaxy is greater 
than about 10\% of the mass of the group and its orbit is highly 
non-radial, 
the analytic model reproduces the mass loss in the simulations 
to $\approx 10\%$ accuracy at all times and for all the orbits, 
galaxy masses, and galaxy concentrations that we have explored.  
For cases where the mass of the galaxy exceeds 10\% of the mass 
of the group, it will likely be necessary to factor in the 
effects of tidal stripping and gravitational shock heating, which 
are neglected by our model.

We re-iterate that the numerical simulations with which our 
analytic model has been calibrated have been demonstrated to be 
robust to the adopted resolution and artificial viscosity 
strength (see \S 2.2).  Furthermore, as we have demonstrated that 
KH (and RT) instability stripping is expected to be unimportant, 
SPH codes should be fully capable of tackling the problem of 
hot halo gas stripping in galaxies orbiting in groups 
and clusters.  A direct comparison between the results using the 
GADGET-2 and FLASH hydrodynamic codes for one of our runs (see 
Appendix A) confirms this conclusion.

The model we have derived has a number of potentially interesting 
applications, including modelling observed satellite galaxies 
and satellite galaxies in cosmological simulations.  One 
application that we are currently pursuing is the incorporation 
of our ram pressure stripping model into a semi-analytic model of 
galaxy formation.  As mentioned in \S 1, recent observations 
(Weinmann et al.\ 2006; Baldry et al.\ 2006) have revealed 
that current semi-analytic models predict 
satellite galaxies whose colours are too red compared to the 
observed systems.  The implementation of ram pressure stripping 
in these models is unrealistically efficient since, by 
assumption, 
the hot halo of the satellite galaxy is instantly transferred to 
the more massive system as soon as the satellite galaxy enters 
the massive system's virial radius.  In reality, the hot gaseous 
halo of the galaxy will remain intact for a while.  For example, 
for the most common orbital parameters, we find that between 
20\%-40\% of the initial hot halo of the galaxy can remain in 
place even after 10 Gyr of orbiting inside the group or cluster 
(see Fig.\ 9; note, however, that the quoted numbers could 
be sensitive to the adopted hot gas distribution of the galaxy).
We note that these predictions are in qualitative agreement with 
recent {\it Chandra} X-ray observations of massive galaxies 
orbiting in hot clusters by Sun et al.\ (2007b), who find that 
most of the galaxies have detectable hot gaseous halos.  
Depending on the efficiency of feedback (e.g., from supernovae 
winds) in the semi-analytic models, radiative cooling of the 
remaining hot halo gas will replenish the cold gaseous component 
at the centre of the galaxy, which in turn will allow star 
formation to continue for some time.  This will have the effect 
of making the colour of model satellite galaxies bluer and could 
resolve the discrepancy between semi-analytic models and 
observations (Font et al., in prep).

\section*{Acknowledgments}

The authors thank the referee for useful suggestions that 
improved the paper and they thank Simone Weinmann, Andrew 
Benson, Volker Springel and Frank van den Bosch for helpful 
discussions.  IGM 
acknowledges support from a NSERC Postdoctoral Fellowship.  CSF 
acknowledges a Royal Society Wolfson Research Merit Award.  ASF 
acknowledges support from a PPARC Postdoctoral Fellowship.  RGB 
acknowledges support from a PPARC Senior Fellowship. MLB 
acknowledges support from a NSERC Discovery Grant. This work was 
supported in part by a PPARC rolling grant to Durham 
University.  Some of the software used in this work was in part 
developed by the DOE-supported ASC / Alliance Center for 
Astrophysical Thermonuclear Flashes at the University of 
Chicago.

\section*{APPENDIX A: COMPARISON OF RAM PRESSURE STRIPPING USING 
GADGET-2 AND FLASH}

Here, we compare the results obtained using the Lagrangian 
SPH code GADGET-2 (Springel 2005) with those obtained using the 
Eulerian AMR code FLASH (Fryxell et al.\ 2000) for one of the 
uniform medium runs (specifically, the run presented in the 
top panel of Fig.\ 4).

The characteristics of the GADGET-2 simulation are given in 
\S 2.1 and \S 3.2 of the main text.  For FLASH, we have tried 
three different versions of the same uniform medium run.  In 
the first version, the galaxy is moved across a periodic box 
filled with a static background medium (as in the GADGET-2 
simulation) and the computational volume is resolved 
with a fixed $256^3$ base grid.  This yields a spatial resolution 
comparable to that of the GADGET-2 run.  In the second version, 
we take advantage of the AMR capability of FLASH, using a base 
grid of $64^3$ cells and allowing up to two levels of refinement.  
This significantly speeds up the 
calculation.  Finally, the third version is the same as the 
first version except that the galaxy is placed in the centre of 
the 
box and is assigned zero bulk velocity while the uniform 
background medium is assigned a velocity of $-1000$ km 
s$^{-1}$.  Encouragingly, we find that all three versions of the 
FLASH run yield virtually identical results.  Below, we compare 
only the results of the first version with the results of the 
GADGET-2 simulation.

\begin{figure}
\centering
\includegraphics[width=8.4cm]{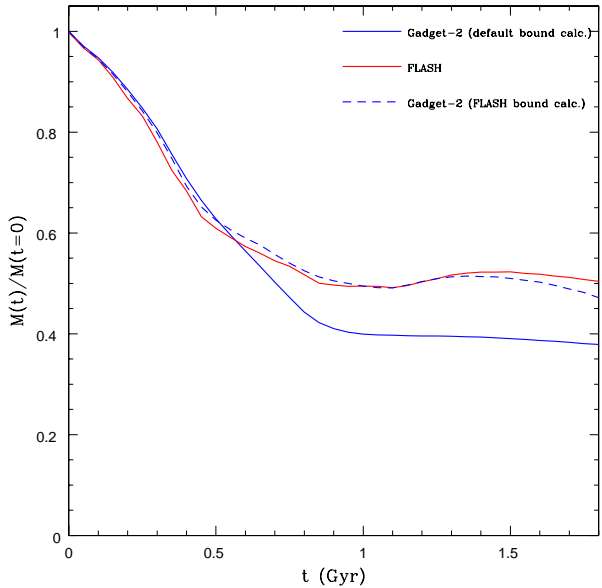}
\caption{A comparison of the GADGET-2 and FLASH results for
the bound mass of gas for the uniform medium run presented in
the top panel of Fig.\ 4.  The solid blue curve corresponds to
applying the default iterative bound mass algorithm described 
at the end of \S 2.1 to the GADGET-2 run.  The solid red curve 
are results of the FLASH code.  The dashed blue curve
corresponds to the case when we apply the same bound mass 
algorithm used for the FLASH run (see text) to the GADGET-2 run.  
This demonstrates that when the GADGET-2 and FLASH runs are 
treated on an equal footing the agreement between the two is 
excellent.
 }
\end{figure}

For all of the GADGET-2 simulations presented in the main text, 
the bound mass of gas is determined by calculating the centre 
of mass of the galaxy, computing energies in this frame, 
throwing out unbound particles, recomputing the centre of mass, 
and so on until no further particles are identified as being 
unbound.  Under this iterative scenario, once a particle is 
stripped it can never be re-accreted.  The $M(t)$ curves are 
necessarily monotonically decreasing in this case.  
Unfortunately, it is not trivial to implement this type of 
algorithm for the FLASH simulation since it is not a Lagrangian 
code.  Instead, it is simply assumed that all of the dark matter 
remains gravitationally bound (this is a good assumption, as 
indicated by the dashed blue curve in Fig.\ 4).  We then use 
the dark matter halo to 
calculate the centre of mass of the galaxy and determine which 
of the gas cells in the box are gravitationally bound to this 
dark matter halo.  Under this scenario, gas that was once 
stripped can potentially be re-accreted.  Therefore, a direct 
comparison between the default GADGET-2 result and FLASH 
result should be treated with caution.  Fortunately, however, 
it is straightforward to apply the same simplified bound mass 
algorithm used for the FLASH run to the GADGET-2 run and we have 
done this.

In Fig.\ 12, we compare the bound mass of gas as a function of 
time for the GADGET-2 and FLASH runs.  The plot demonstrates 
that when both runs are treated on an equal footing, using the 
same algorithm for computing the bound mass of gas, the 
agreement between them is superb.  


\section*{APPENDIX B: THE IMPORTANCE OF TIDAL STRIPPING}

Here, we present a simple argument that demonstrates that tidal 
stripping should only be relevant for cases where the mass of the 
galaxy exceeds $\sim 10\%$ of the mass of the group.

The tidal radius, $r_t$, of a galaxy can be expressed as

\begin{equation}
\frac{r_t}{R} = \biggl(\frac{M_{\rm gal}(r_t)}{M_{\rm grp}(R)(3 - 
d\ln{M_{\rm grp}}/d\ln{R})} \biggr)^{1/3} \ \ ,
\end{equation}

\noindent where $R$ is 3D group-centric radius of the galaxy, 
$M_{\rm gal}(r)$ is the total mass of the galaxy within radius 
$r$, and $M_{\rm grp}(R)$ is the total mass of the group within 
radius $R$ (e.g., King 1962).

The above equation can be re-written in terms of the mean density 
of the galaxy within $r_t$ and the mean density of the group 
within $R$:

\begin{equation}
\overline{\rho_{\rm gal}(r_t)} = \biggl(3-\frac{d\ln{M_{\rm 
grp}}}{d\ln{R}} \biggr) \ \overline{\rho_{grp}(R)} \ \ .
\end{equation}

For simplicity, we will now assume that both systems can be 
approximated as isothermal spheres.  In this case, the condition 
for tidal stripping is simply

\begin{equation}
\overline{\rho_{\rm gal}(r_t)} < 2 \ \overline{\rho_{grp}(R)} \ \ 
.
\end{equation}

We now seek to express the ram pressure stripping condition in 
terms of $\overline{\rho_{\rm gal}(r_t)}$ and 
$\overline{\rho_{grp}(R)}$.  

Assuming for both the galaxy and the group that the gas density 
traces the total density and that both have the same baryon 
fraction, eqn.\ (7) can be re-written as

\begin{equation}
\rho_{grp}(R) v_{\rm orb}^2 > \alpha \rho_{\rm gal}(r) v_{\rm c, 
gal}^2(r) \ \ .
\end{equation}

Rearranging, we obtain

\begin{equation}
\rho_{\rm gal}(r) < \frac{1}{\alpha} \biggl(\frac{v_{\rm 
orb}}{v_{\rm c,gal}} \biggr)^2 \rho_{\rm grp}(R) \ \ .
\end{equation}

If both the galaxy and group have the same power law density 
profiles, then

\begin{eqnarray}
\rho_{\rm gal}(r) = k \overline{\rho_{\rm gal}(r)}\\
\rho_{\rm grp}(r) = k \overline{\rho_{\rm grp}(r)} \nonumber
\end{eqnarray}

\noindent for some $k$.

Therefore, the ram pressure stripping condition is given by

\begin{equation}
\overline{\rho_{\rm gal}(r)} < \frac{1}{\alpha} 
\biggl(\frac{v_{\rm
orb}}{v_{\rm c,gal}} \biggr)^2 \ \overline{\rho_{\rm grp}(R)} \ \ 
,
\end{equation}

\noindent which is similar to the tidal stripping condition 
(eqn.\ 11) except that the right-hand side is larger by a factor 
$F$:

\begin{equation}
F = \frac{1}{2 \alpha} \biggl(\frac{v_{\rm
orb}}{v_{\rm c,gal}} \biggr)^2 \ \ .
\end{equation}

Typically, $v_{\rm orb} \sim v_{\rm c,grp}$ (where $v_{\rm 
c,grp}$ is the circular of the group) and assuming $\alpha = 2$ 
the factor $F$ can be expressed as

\begin{equation}
F \sim \frac{1}{4} \biggl(\frac{v_{\rm
c,grp}}{v_{\rm c,gal}} \biggr)^2 \ \ .
\end{equation}

Since $v_c \propto M^{1/3}$ for cosmological halos, eqn.\ (17) 
can be re-written as

\begin{equation}
F \sim \frac{1}{4} \biggl(\frac{M_{\rm grp}}{M_{\rm gal}} 
\biggr)^{2/3} \ \ .
\end{equation}

Tidal stripping is therefore only expected to become more 
important than ram pressure stripping (i.e., $F \ga 1$) in cases 
where $M_{\rm gal}/M_{\rm grp} \ga 1/8$.

\bsp

\label{lastpage}

\end{document}